# Characterization of transient discharges under atmospheric-pressure conditions applying nitrogen photoemission and current measurements

**Sandra Keller[1,2], Priyadarshini Rajasekaran[2], Nikita Bibinov[2] and Peter Awakowicz[2]**
[1]ERBE Elektromedizin GmbH, Waldhoernlestrasse 17, 72072 Tuebingen, Germany
[2]Institute for Electrical Engineering and Plasma Technology, Ruhr-Universitaet Bochum, 44801 Bochum, Germany
E-mail: sandra.keller@erbe-med.com,
  rajasekaran@aept.ruhr-uni-bochum.de,
  nikita.bibinov@rub.de,
  awakowicz@aept.rub.de

**Abstract.** The plasma parameters such as electron distribution function and electron density of three atmospheric-pressure transient discharges namely filamentary and homogeneous dielectric barrier discharges in air, and the spark discharge of argon plasma coagulation (APC) system are determined. A combination of numerical simulation as well as diagnostic methods including current measurement and optical emission spectroscopy (OES) based on nitrogen emissions is used. The applied methods supplement each other and resolve problems, which arise when these methods are used individually. Nitrogen is used as sensor gas and is admixed in low amount to argon for characterizing the APC discharge. Both direct and stepwise electron impact excitation of nitrogen emissions are included in the plasma-chemical model applied for characterization of these transient discharges using OES where ambiguity arises in the determination of plasma parameters at specific discharge conditions. It is shown that the measured current solves this problem by providing additional information useful for the determination of discharge-specific plasma parameters.

# 1. Introduction

Atmospheric pressure discharges have wide field of applications in industry and medicine as sources for generation of chemical active species like ozone, for treatment of living organisms, etc. The densities of charged particles in the discharges at atmospheric-pressure conditions are usually high that causes high gas temperature. To prevent overheating of the treated object in medical applications or during treatment of thermolabile materials, these discharges are operated in the transient mode with plasma duration in the range of several nanoseconds to a few microseconds. For this purpose, pulsed power supplies are applied and a dielectric layer can be placed between the electrodes.

Determination of plasma conditions (like gas temperature) and plasma parameters (electron distribution function and electron density) are required for effective optimisation, and thereby the safe application of the plasma sources. Characterization of the plasma sources operated at atmospheric pressure is a complicated task because of high densities of gas and charged species as far as small plasma volumes are concerned. Voltage-current measurements, optical emission spectroscopy (OES), and numerical simulation are applied for characterization of transient discharges operated at atmospheric-pressure conditions [1], [2], [3].

Complicated plasma-chemical models and different assumptions have to be used when these diagnostic methods are applied separately. Combination of experimental and theoretical diagnostic methods like numerical solution of Boltzmann equation and optical emission spectroscopy (OES) are effectively applied for plasma characterization of discharges operated at atmospheric pressure, namely dielectric barrier discharges (DBD) in air [1], microwave discharge in $N_2/O_2$ mixture [4], and radio frequency µ-jet in $He/O_2/N_2$ mixture [5]. Nitrogen acts as sensor gas and OES diagnostics was applied in assumption that all nitrogen molecular emissions are excited during electron impact of ground state $N_2(X)$. However, at some plasma conditions, especially at high electron density, this assumption is not sufficient and a plasma-chemical model, where emission of nitrogen ions also excited by electron impact of ground state of $N_2^+$, must be applied [6]. Therefore, consideration of plasma chemical models of excitation of both nitrogen emissions is necessary.

In certain discharge configuration, e.g. in filamentary discharges or homogeneous discharges in the gap between flat electrodes, current and microphotography measurements can be applied for current density determination. This gives additional information about plasma parameters because current density is a function of electric field and electron density.

In this study, we present the determination of plasma parameters in atmospheric pressure transient discharges taking into account the direct and stepwise excitations of nitrogen molecular emissions. In addition, we show that the reliability of the determined parameters can be ensured by using the measured current in the plasma source as an additional diagnostic tool. The three transient discharges reported here are the homogeneous DBD [3] and the single-filamentary DBD [2] in air (78% nitrogen/ 21% oxygen) as far as the spark discharge ignited in a medical argon plasma coagulation system (APC) [7] in a mixture of 95% argon and 5% nitrogen.

We report the combination of theoretical and experimental methods to determine plasma parameters in these transient discharges that are characterized with distinct plasma conditions.

**2. Diagnostics and plasma characterization**

Nitrogen is used as sensor gas in our experiments for OES diagnostics. UV-emission of nitrogen ($N_2$(C-B) - second positive system and $N_2^+$(B-X) - first negative system) are excited in the plasma volume. The gas temperature ($T_g$ in K) and the plasma parameters (electron velocity distribution function - $f_v(E)$ in eV$^{-3/2}$, and electron density - $n_e$ in m$^{-3}$) are determined using these emissions. The averaged gas temperature in the active plasma volume is determined using rotational distribution of vibrational band $N_2$(C-B,0-0) at 337.1 nm. Due to fast rotational relaxation at atmospheric-pressure, the rotational temperature ($T_{rot}$) of diatomic molecules is equal to the gas temperature. The measured emission spectrum of $N_2$(C-B,0-0) band is compared with simulated spectra for various rotational temperatures. By fitting, the averaged gas temperature in the active plasma volume is determined [1].

The emission spectrum of plasma ($I_\lambda(\lambda)$ in photon·s$^{-1}$·nm$^{-1}$) is measured using an Echelle spectrometer (ESA 3000, LLA Instruments, Germany) (figure 1) with a spectral resolution $R = \frac{\lambda}{\Delta\lambda} = 13333$ for the wavelength region of λ = 200 - 800 nm corresponding to the FWHM (*full width at half maximum*) of the apparatus-function of Δλ = 0.015 - 0.06 nm. Optic fibre of spectrometer is provided with diaphragm for limitation of the acceptance angle and increase of space resolution by OES diagnostic. The spectrometer is relatively and absolutely calibrated using a tungsten-ribbon lamp and the well-known photoemission of nitric oxide and $N_2$ [8].

The diameter $d_p$ (in m) of the cylindrical microdischarge channel in the filamentary DBD and APC discharge is determined by microphotography using a high-speed sensitive CCD camera (Sensicam qe, PCO Germany). At that uniform radial distribution of photoemission in the plasma channel is assumed and two dimensional Abel transform ($I(r) = C\sqrt{(r_0^2 - r^2)}$, I(r) - measured irradiance, $r_0$ - radius of the plasma channel) of the image is applied.

The measured radial distributions of filamentary DBD and APC discharge (figure 2) slightly differ from the homogenous ones. To determine the diameter of the plasma channel, we calculate the diameter of homogenous plasma channel that provides the equal integral of the intensity in radial direction of the image measured by CCD.

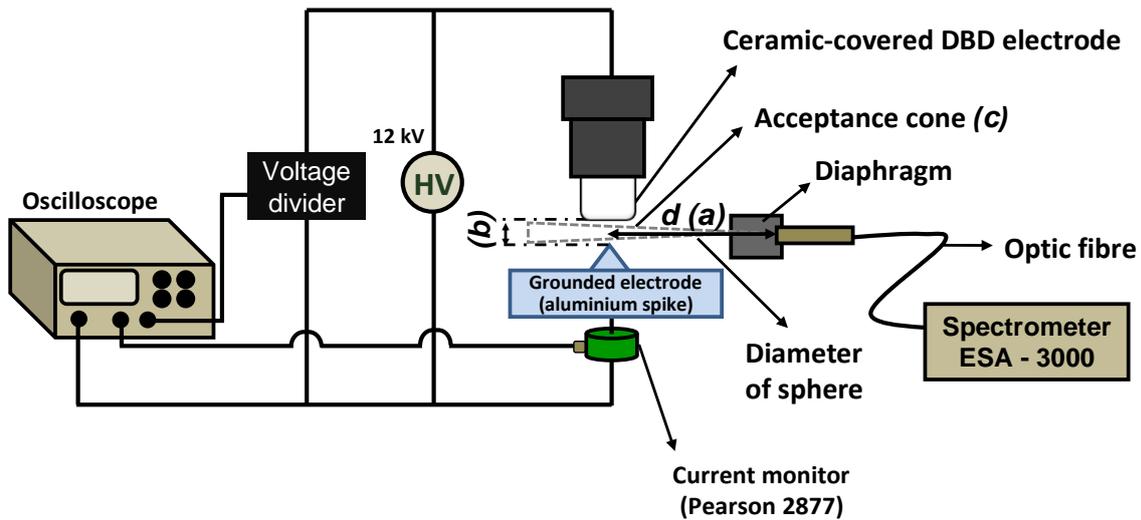

**Figure 1.** Experimental setup by characterization of filamentary DBD using OES. Similar setup is applied by characterization of homogeneous DBD and APC discharge. (a) d – distance between the tip of the spike and the entrance hole of the optic fibre; (b) distance (= 3.4 mm) between the spike and ceramic covering of the DBD electrode ; (c) acceptance cone is half than interelectrode distance

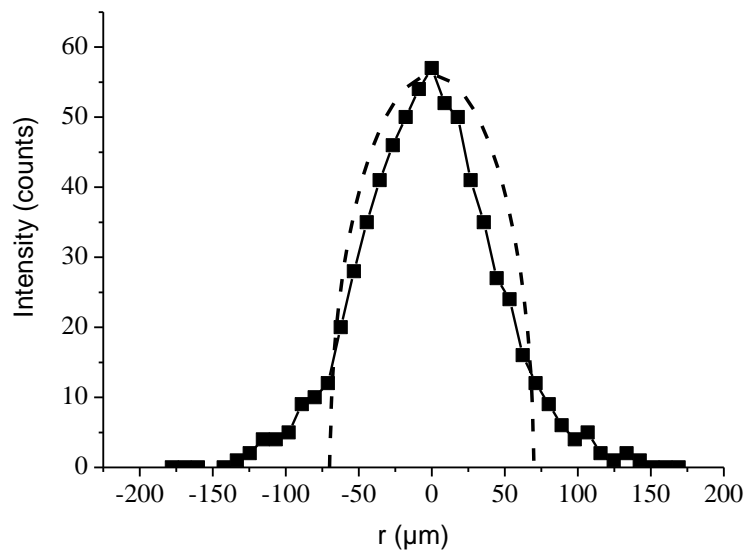

**Figure 2.** Radial intensity distribution in APC discharge image determined by CCD (■). Intensity distribution in image of uniform plasma channel with equal integral in radial direction (dotted line)

For homogeneous DBD, the diameter is equal to diameter of powered electrode, because the whole gap between the electrodes is filled with plasma.

The electric current and the plasma duration are determined from the current pulse measured using a current monitor (Pearson$^{TM}$ Current Monitor – 2877 with 1 V: 1 A output). The current and voltage traces are recorded using an oscilloscope (LeCroy Waverunner 204 Xi - A, 2 GHz), whereby the voltage is measured using a voltage divider (figure 1).

The intensities of molecular bands $I_{N_2(C-B,0-0)}$ and $I_{N_2^+(B-X,0-0)}$ (in photon·s$^{-1}$·m$^{-3}$) are obtained by integrating the measured spectrum corrected to the efficiency of the spectrometer (1) and divided to observed plasma volume ($V_p^{obs}$ in m$^{-3}$) and geometrical factor (G)

$$I_{N_2(C-B,0-0)} = \frac{\int_{\lambda_1}^{\lambda_2} I_\lambda d\lambda}{V_p^{obs} \cdot G} \quad (1)$$

For calculation of the plasma volume observed by the spectrometer, the acceptance angle of the spectrometer obtained with the fitted diaphragm and the geometrical factor are taken into account. The geometrical factor is equal to the ratio between the circular area of the entrance hole of the optic fibre and the surface area of the sphere with radius *d* (in m) - the distance between the optic fibre and the active plasma volume (see figure 1).

In the case of single-filamentary DBD, the averaged plasma parameters in the channel can be effectively influenced by the large electric field due to the surface discharge on the ceramic [1] as well as on the tip of the spike. Hence, the optic fibre was positioned in such a way that the emission from the "body" of the microdischarge channel between the electrodes is observed. The height of the microdischarge channel seen by the spectrometer during characterization of filamentary DBD is 1.6 mm including the acceptance cone (figure 1).

At our experimental conditions, nitrogen photoemissions N$_2$(C-B) and N$_2^+$(B-X) can be excited by electron impact and by collisions with argon metastables ((2-10) and figure 3).

$$N_2(X) + e \rightarrow N_2(C) + e \quad (2)$$
$$N_2(X) + e \rightarrow N_2^+(B) + 2e \quad (3)$$
$$N_2^+(X) + e \rightarrow N_2^+(B) + e \quad (4)$$
$$N_2(X) + e \rightarrow N_2(A) + e \quad (5)$$
$$N_2(A) + e \rightarrow N_2(B) + e \quad (6)$$
$$N_2(A) + e \rightarrow N_2(C) + e \quad (7)$$
$$N_2(A) + e \rightarrow N_2^+ + 2e \quad (8)$$
$$Ar + e \rightarrow Ar^* + e \quad (9)$$
$$Ar^* + N_2(X) \rightarrow N_2(C) + Ar \quad (10)$$

N$_2$(C-B,0-0) can be excited by direct electron impact excitation of ground state neutral nitrogen molecule N$_2$(X) (2) as well as by stepwise excitation via neutral metastable state N$_2$(A) (5) and by collisions with excited argon atoms (10). The N$_2^+$(B-X,0-0) emission can be excited by electron impact of N$_2$(X) (3) and ground state molecular ion N$_2^+$(X) (4), and both excitation processes are considered in the presented study (figure 3). Processes (9, 10) are not considered in the case of DBD because of low concentration of argon in air plasma.

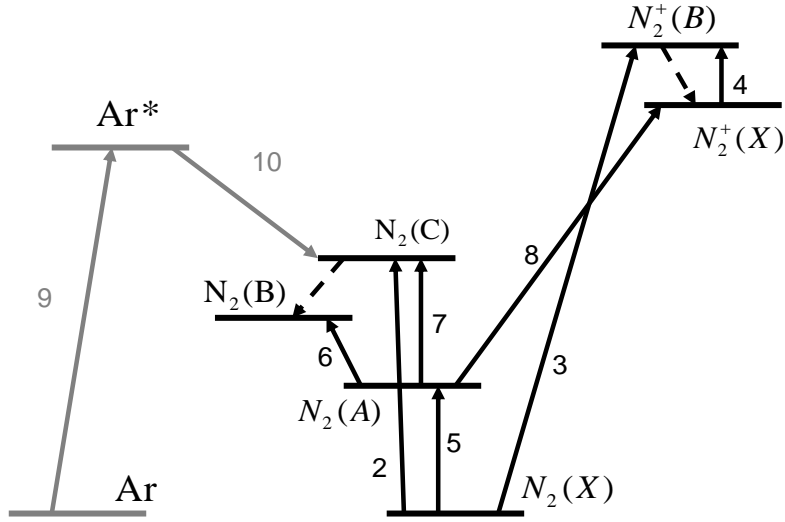

**Figure 3.** General scheme of energy levels and processes considered for OES diagnostic applying $N_2(C-B)$ and $N_2^+(B-X)$ emissions by characterization of DBD and APC discharge. Processes (9, 10) are not considered in the case of DBD because of low concentration of argon in air plasma.

The ratio of intensities of nitrogen emission bands in the DBD and APC discharge depends on the electron distribution function and the density of nitrogen molecular ions (11):

$$\frac{I_{N_2^+(B-X,0-0)}}{I_{N_2(C-B,0-0)}} = \frac{Q_{N_2^+(B)} \cdot \left(N_{N_2} \cdot k_{N_2^+(B)} + N_{N_2^+(X)} \cdot k_{N_2^+(B)}^{N_2^+}\right) \cdot n_e}{Q_{N_2(C)} \cdot N_{N_2} \cdot n_e \cdot \left(k_{N_2(C)} + K_{N_2(C)}^{Ar^*} + K_{N_2(C)}^{N_2(A)}\right)} \quad (11)$$

where, $Q_{N_2^+(B)} = \dfrac{A'}{A' + \sum\limits_M k_{qM}^{N_2^+(B)} \cdot N_M}$ and $Q_{N_2(C)} = \dfrac{A}{A + \sum\limits_M k_{qM}^{N_2(C)} \cdot N_M}$, $A'$ and $A$ (in $s^{-1}$) are the

corresponding Einstein coefficients [9], [10], $k_{qM}^{N_2^+(B)}$ (in $m^3 \cdot s^{-1}$) is the rate constant for quenching of $N_2^+(B)$ excited states during collision with gas species $M$ ($N_2$ and $O_2$ for DBD [8], and $N_2$ and Ar for APC discharge [9]), $N_{N_2^+(X)}$ (in $m^{-3}$) is the density of nitrogen molecular ions. $k_{N_2^+(B)}$ and $k_{N_2(C)}$ (in $m^3 \cdot s^{-1}$) are the rate constants, respectively, for photoemission from level $N_2(C)$ to $N_2(B)$ (*abbreviated $N_2(C-B)$*) and $N_2^+(B)$ to $N_2^+(X)$ (*abbreviated $N_2^+(B-X)$*), where the levels $N_2(C)$ and $N_2^+(B)$ are populated by electron impact excitation from $N_2(X)$. $k_{N_2^+(B)}^{N_2^+}$ is the rate constant for electron impact excitation of $N_2^+(B-X)$ emission from $N_2^+(X)$. In (11), $K_{N_2(C)}^{Ar^*} = \dfrac{N_{Ar} \cdot k_{N_2(C)}^{Ar^*} \cdot k_{Ar^*} \cdot (B1) \cdot (B2)}{n_e \cdot k_{qAr^*}^e + N_{Ar}^2 \cdot k_{qAr^*}^{2Ar} + N_{N_2} \cdot k_{qAr^*}^{N_2}}$ is included only for APC discharge, where $k_{N_2(C)}^{Ar^*}$ is the rate constant for the excitation of N$_2$(C) emission by collisions of argon metastables with nitrogen (2.8·10$^{-17}$ m$^3$·s$^{-1}$ at 300 K [9]), $k_{Ar^*}$ is the rate constant for electron impact excitation of argon metastables and resonance states (in m$^3$·s$^{-1}$), B1 = 0.787 and B2 = 0.5 are the branching in vibrational distributions of N$_2$(C-B) by excitation of nitrogen in collisions with excited argon atoms [9], $k_{qAr^*}^e$, $k_{qAr^*}^{2Ar}$, $k_{qAr^*}^{N_2}$ are the rate constants of corresponding processes for quenching of excited Ar by collisions with electrons (ionisation), two argon atoms and nitrogen molecule ($k_{qAr^*}^{2Ar}$ = 1.2·10$^{-44}$ m$^6$·s$^{-1}$ [9], $k_{qAr^*}^{N_2}$ = 3.5·10$^{-17}$ m$^3$·s$^{-1}$ [9]), and $N_{Ar}$ is the density of Ar (in m$^{-3}$). Because of the similarity of energy terms of metastable and resonance argon states, and due to trapping of photoemission at atmospheric conditions, we include the resonance state with the rate constants determined for metastable states [10] by consideration of simulation of nitrogen excitation. $K_{N_2(C)}^{N_2(A)}$ in equation (11) is the rate constant for production of N$_2$(C) from N$_2$(A), determined by $K_{N_2(C)}^{N_2(A)} = \dfrac{k_{N_2(A)}^{N_2(X)} \cdot k_{N_2(C)}^{N_2(A)} \cdot (B2) \cdot (B3)}{k_{N_2(B)}^{N_2(A)} + k_{N_2(C)}^{N_2(A)} + k_{N_2^+}^{N_2(A)}}$. Where, $k_{N_2(A)}^{N_2(X)}$ (in m³·s$^{-1}$) is the rate constant for electron impact excitation of N$_2$(A) state from N$_2$(X), $k_{N_2(C)}^{N_2(A)}$, $k_{N_2(B)}^{N_2(A)}$, and $k_{N_2^+}^{N_2(A)}$ (in m³·s$^{-1}$) are the rate constants for electron impact excitation of N$_2$(C), N$_2$(B), and N$_2^+$ from N$_2$(A). B3 is branching in vibrational distributions of N$_2$(C) (B3 = 0.5) that is determined in our experiment using vibrational distribution in emission of nitrogen and known Frank-Condon factors [11].

The excitation rate constants by electron impact depend on $f_v(E)$ and the excitation cross-section $\sigma_{exc}$ (in m$^2$) [12], [13], [14], [15], [16] as shown in (12):

$$k_{exc} = 4\pi\sqrt{2} \int_0^\infty f_v(E) \sqrt{\dfrac{2e}{m} E} \cdot \sigma_{exc}(E)\, dE, \qquad (12)$$

where e and m are elementary charge (in C) and mass of an electron (in kg). Here, E is the kinetic energy of electrons (in eV). $f_v(E)$ is normalized to fulfil (13):

$$4\pi\sqrt{2} \int_0^\infty f_v(E) \sqrt{E}\, dE = 1 \qquad (13)$$

To calculate these rate constants, we solve the Boltzmann equation (14) numerically in 'local approximation' for discharge-specific gas mixtures (DBD: nitrogen/ oxygen = 78%/ 21%; APC discharge: argon/ nitrogen = 95%/ 5%) [17]. The program code 'EEDF' developed by A P Napartovich et al. [18] is used for this purpose,

$$\dfrac{\partial f}{\partial t} - \left(\dfrac{e}{m}\vec{E}\right) \cdot \nabla_v f = \left.\dfrac{\partial f}{\partial t}\right|_{collisions} \qquad (14)$$

where f is the electron distribution function, $\nabla_V$ is the gradient in velocity space and $\vec{E}$ is the electric field strength. The term on the right hand side of equation (14) includes all kinds of binary electron-heavy particle collisions. Electric field at our experimental conditions is not known *a priory*, hence we simulate $f_v(E)$ in a broad range of electric field strength values (14) and calculate excitation rate constants by electron impact using (12), which are used in plasma-chemical modelling.

Applying rate constants for ionisation of working gas species and lifetime of positive ions at DBD and APC discharge conditions, we determine the density of nitrogen ions $N_{N_2^+(X)}$ in assumption of quasi-neutrality (15):

$$N_{N_2^+(X)} = n_e \cdot \frac{N_{N_2^+}}{N_{O_2^+} + N_{N_2^+} + N_{Ar^+} + N_{Ar_2^+}} = n_e \cdot R_i \qquad (15)$$

After substitution of $N_{N_2^+(X)}$, equation (11) for intensity distribution in the emission spectrum of nitrogen can be used in a more general case for the determination of electron density, which is a function of the reduced electric field (16):

$$n_e = \frac{N_{N_2} \cdot \left( \frac{I_{N_2^+(B-X,0-0)} \cdot Q_{N_2(C)}}{I_{N_2(C-B,0-0)} \cdot Q_{N_2^+(B)}} \cdot \left( k_{N_2(C)} + K_{N_2(C)}^{Ar^*} + K_{N_2(C)}^{N_2(A)} \right) - k_{N_2^+(B)} \right)}{R_i \cdot k_{N_2^+(B)}^{N_2^+}} = F(E/N) \qquad (16)$$

Absolute intensity of nitrogen emission $I_{N_2(C-B,0-0)}$ and electric current are functions of electric field and electron density and also used by determination of plasma parameters (17, 18)

$$n_e = \frac{I_{N_2(C-B,0-0)}}{Q_{N_2(C)} \cdot N_{N_2} \cdot \left( k_{N_2(C)} + K_{N_2(C)}^{Ar^*} + K_{N_2(C)}^{N_2(A)} \right)} = F(E/N) \qquad (17)$$

$$n_e = \frac{j}{e \cdot v_d} = F(E/N) \qquad (18)$$

where, j is the current density (in A·m$^{-2}$), and $v_d$ is drift velocity of electrons (in m·s$^{-1}$) at corresponding plasma conditions calculated applying the program code 'EEDF'.

The plasma parameters (electric field and electron density) at our experimental conditions are determined by solution of a system of equations (16-18). Equations (16-18) present a system of non-linear equations which are used for the determination of electron density, and reduced electric field and the corresponding electron distribution function. As discussed earlier, $K_{N_2(C)}^{Ar^*}$ is a function of electron density and also has an influence on the electron density determined using (16) and (17). Hence, we determine $n_e$ in APC discharge operated in Ar/N$_2$ mixture applying (16) and (17) and method of successive approximations (see below).

As was mentioned above, by determination of averaged plasma parameters for DBD operated in air the density of argon is negligible. Because of the high velocity of the working gas and the small distance between powered electrode and tissue the influence of oxygen on plasma-chemical conditions in APC discharge operated in an Ar/N$_2$ mixture is also negligible. The validation of this assumption is tested by application of a quartz tube placed around the powered electrode, which protects the discharge from the influence of surrounding air. No differences in emission spectra of nitrogen and argon are established by application of this quartz tube.

The rates of direct and stepwise excitations of nitrogen photoemissions are proportional to steady state densities of the species, namely nitrogen molecule's N$_2$(X) and nitrogen molecular ion's N$_2^+$(X) in ground state, and nitrogen metastable's N$_2$(A), as far as the rate constants for the corresponding electron impact excitations. At considered plasma conditions, the density of nitrogen molecules in ground state can be calculated using gas temperature, pressure, and working gas mixture. The density of nitrogen molecular ions is determined in our study using electron density in assumption of plasma quasi-neutrality and partial density of nitrogen molecular ions that is function of corresponding ionization rates, and lifetime of molecular and atomic positive ions at considered plasma conditions.

The lifetime and steady state concentration of metastable N$_2$(A) depends on the rate constants of quenching and the densities of the quenchers. At atmospheric pressure plasma conditions in air, the nitrogen metastables are quenched effectively in collisions with oxygen molecules and atoms, ozone and nitrogen oxides [19]. Because of this reason the density of nitrogen metastable N$_2$(A) at atmospheric pressure conditions in air amounts about $10^{20}$ m$^{-3}$ [20], which is about five orders of magnitude lower than the density of nitrogen in ground state. The rate constants of electron impact excitation of N$_2$(C-B) emission from N$_2$(A) determined using (12) and cross section $\sigma_{exc}(E)$ calculated in [15] are about three orders of magnitude higher than the rate constants determined for direct electron impact excitation from nitrogen ground state. Therefore stepwise excitation of N$_2$(C-B) photoemission excited from nitrogen metastable N$_2$(A) is negligible by characterization of DBD plasma in air.

Plasma parameters of transient discharges are determined in our experiment with low space and time resolution. We determined temporary averaged electric current density in assumption of radial homogeneous distribution and nitrogen emission spectra with resolution of about 1 mm and 1 µs. At that, the spectrometer observes regions of different plasma conditions (electric field and electron density) and the current is not constant during the active phase of the discharge. Such experimental method cannot be used for the study of a temporal and space behavior of plasma parameters in transient discharges in contrast to cases [21] where the maximum value of the electric field near the head of the streamer is determined by using OES after the corresponding corrections. Our diagnostic method is based on averaged OES and current measurements, and can provide important information concerning averaged rates of the production of chemical active radicals or excitation of photoemissions by application of different plasma sources. These can be calculated applying temporary and space averaged plasma parameters. The influence of inhomogeneity of the plasma parameters on calculated excitation (or dissociation) rates by application of OES diagnostics was studied in [9], [22], [6], and a good agreement between space resolved and averaged excitation rates was found in most cases.

## 3. Atmospheric-pressure DBD in air: Discharge modes and plasma characterization
*3.1. The DBD device and discharge modes*

The DBD device [1] consists of a ring-shaped powered copper electrode (diameter = 8 mm) covered with ceramic ($Al_2O_3$) of 1 mm thickness. The counter electrode can be objects of high capacitance or grounded electrodes of different materials and profiles (flat or spike). A high voltage pulsed power supply of 300 Hz trigger frequency is used to ignite a discharge in ambient air. Each trigger pulse initiates a sequence of high voltage oscillations with damped amplitude; the frequency within each sequence is 100 kHz with maximum amplitude of about 12 kV [1]. Different counter electrodes namely metal, glass and liquid are used [3] for plasma characterization. A grounded aluminium spike ignites a single-filamentary discharge (figure 4) [2] while glass, when used as the counter electrode, produces a homogenous discharge [3] for the same applied voltage and trigger frequency. A detailed discussion on the formation of these different modes is already presented [3]. The plasma duration depends on the applied voltage and the inter-electrode distance. In our experiment it is 20 ns for the single-filamentary DBD and 18 ns for the homogeneous DBD. The current and voltage profiles for homogeneous mode are shown in figure 5.

In emission spectrum of DBD in air, the vibrational bands of 'second positive' ($N_2$(C-B)) and 'first negative' ($N_2^+$(B-X)) systems of nitrogen are observed. In figure 6, a typically optical emission spectrum for homogeneous DBD in air is shown. Intensity distribution in vibrational spectrum of nitrogen depends on the known Frank-Condon factors [11], [23] by photoemission and electron impact excitation, and also on vibrational distribution in ground state. The latter can be determined applying measured emission spectrum.

In UV-spectral range of emission spectrum measured for APC discharge not only nitrogen bands are observed but also other diatomic molecules like NH(A-X) and CN(B-X). The overlapping of these bands with nitrogen emission causes a high inaccuracy by gas temperature measurement and determination of drift velocity of positive ions.

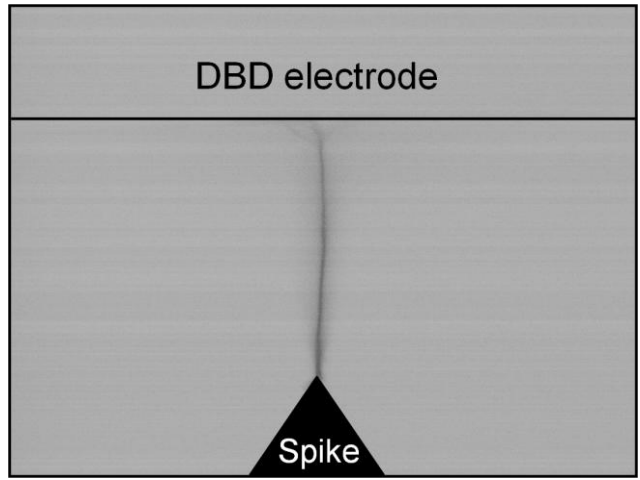

**Figure 4.** Inverse microphotograph of single-filamentary DBD ignited in air with grounded aluminium spike as the opposite electrode. (Only a section of the DBD electrode is shown). Amplitude of applied voltage = 12 kV. Distance between electrodes = 3.4 mm.

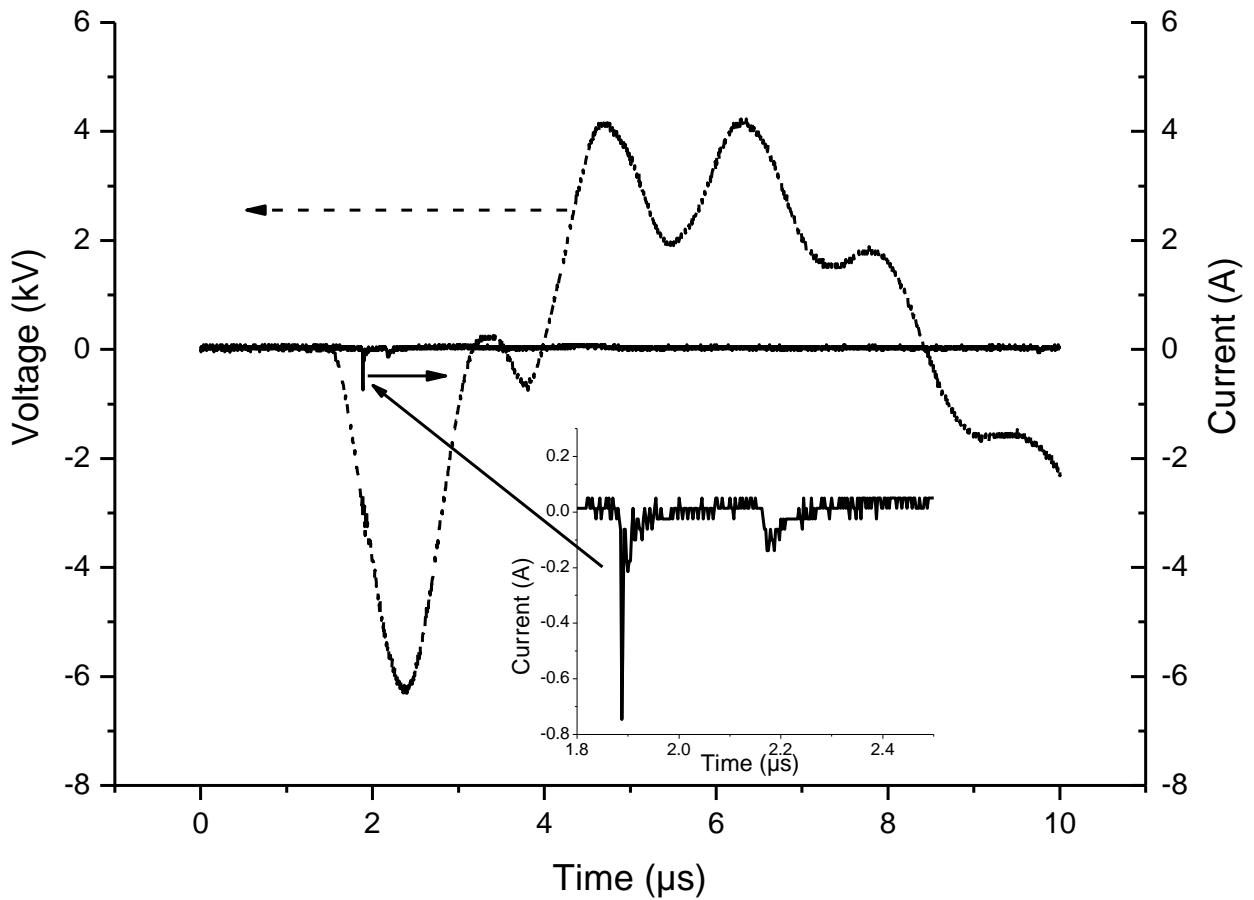

**Figure 5.** Measured current and voltage profiles for homogeneous DBD in air. Averaged duration of current pulse (FWHM) amounts to 18 ns.

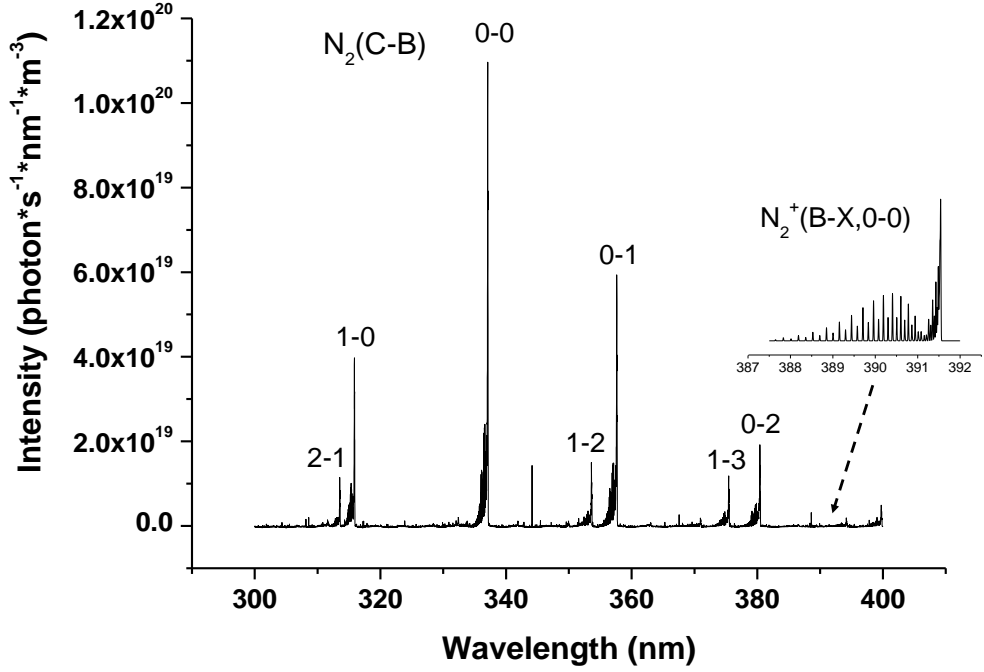

**Figure 6.** Emission spectrum measured for homogeneous DBD in air.

*3.2. Plasma parameters in single-filamentary and homogeneous DBD modes*

For the characterization of DBD, we determine $R_i = \dfrac{N_{N_2^+}}{N_{N_2^+} + N_{O_2^+}}$ in air at atmospheric pressure, where $N_{N_2^+}$ and $N_{O_2^+}$ are the densities of nitrogen and oxygen molecular ions (in m$^{-3}$), calculated by taking into account the ionization and recombination of molecular nitrogen and oxygen ions. Electron-ion recombination rate constant for nitrogen ions is calculated using equation (12) and cross section from [12]. The expression $2 \times 10^{-13} \left( 300K / T_e \right)$ m$^3 \cdot$s$^{-1}$ [24] is applied for determination of the recombination rate constant for electron-oxygen ion recombination. At that, electron temperature is determined by averaging the EVDF in the region of electron kinetic energy of 0 - 10 eV.

We calculate electron densities applying measured intensities of nitrogen emissions and current densities using (16-18) in broad region of electron kinetic energy, and present these dependences in graphical form (see figure 7) to find all possible solutions to the system of equations, and rule out possible ambiguity, if any. The lines corresponding to the ratio of nitrogen intensities determined in assumption of only direct electron impact excitation of nitrogen emissions (19) are also presented in figure 7.

$$\frac{I_{N_2^+(B-X,0-0)}}{I_{N_2(C-B,0-0)}} = \frac{Q_{N_2^+(B)} \cdot N_{N_2} \cdot k_{N_2^+(B)} \cdot n_e}{Q_{N_2(C)} \cdot N_{N_2} \cdot k_{N_2(C)} \cdot n_e} \qquad (19)$$

An exact solution of the system of equations (16-18) is given by the intersection of all three curves in one point. But because of inaccuracies of measured values and applied rate constants, the intersection of these curves can give three points (see figure 7) close to each other. In this case, the reliable solution must be determined after the analysis of the confidence intervals of the measured values and used rate constants. The reliability of the determined plasma parameters will be discussed below. Here, the aim is to emphasize that in both modes (filamentary and homogeneous) of DBD, the averaged electric field is high (about 200 Td). The intersections of the curves in the lower electric field region correspond to the plasma conditions, where emission intensities and current cannot be measured simultaneously and cannot give reliable plasma parameters.

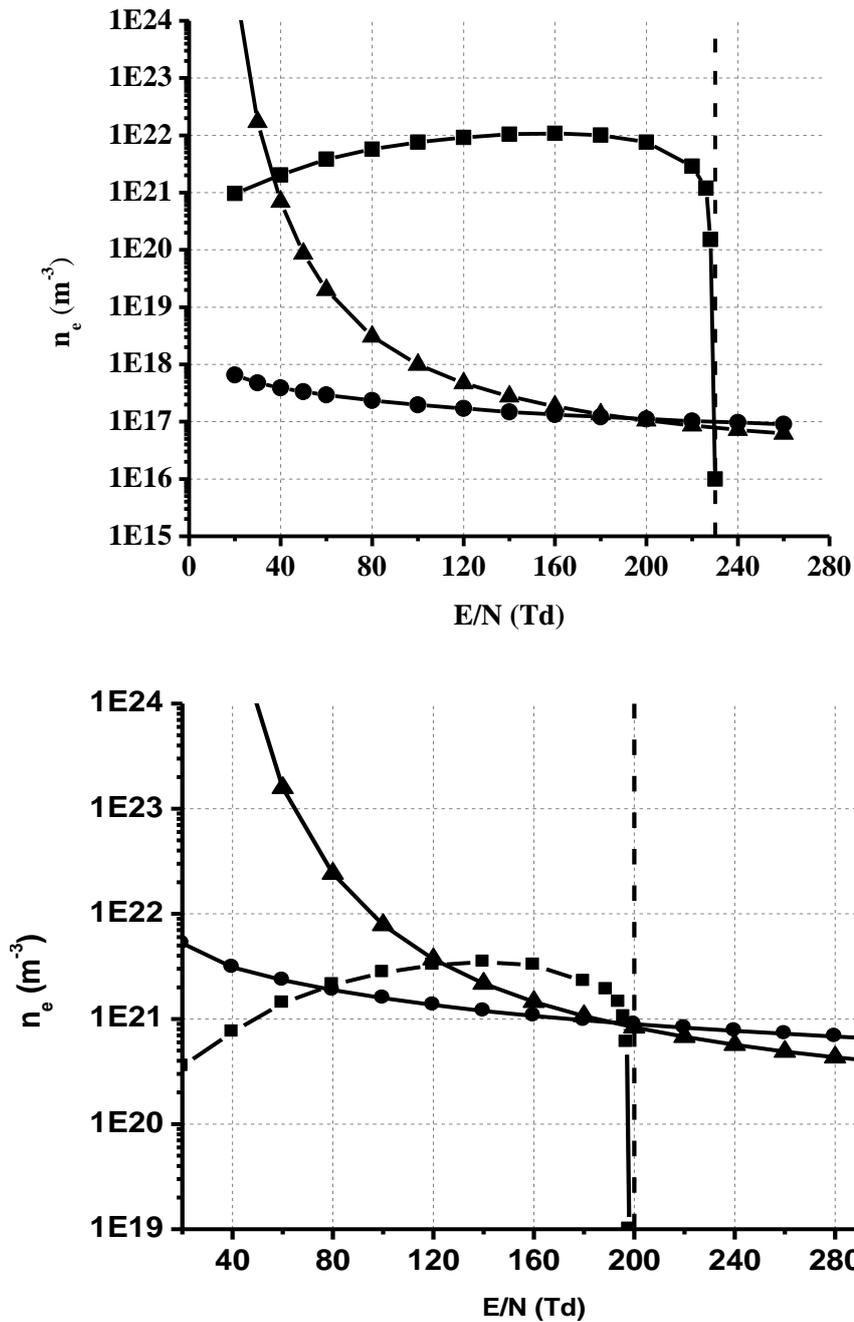

**Figure 7.** Graphical interpretation of equations system (16-18) for homogeneous DBD (top) and single-filamentary DBD (bottom). ■ - applying ratio of nitrogen emissions (16), ▲ - absolute intensity of $N_2$(C-B,0-0) (17), ● - current density (18). Dotted line presents the electric field determined applying ratio of nitrogen intensities in assumption of only direct electron impact excitation (19).

## 4. Argon plasma coagulation (APC) system

The argon plasma coagulation system comprises of an APC unit (ERBE APC 2 system, ERBE Elektromedizin GmbH, Tuebingen, Germany), an electrosurgical generator (ERBE VIO 300 D, ERBE Elektromedizin GmbH, Tuebingen, Germany), and an APC probe (ERBE Elektromedizin GmbH,

Ref.: 20132-156). This consists of a flexible poly-tetrafluoroethylene (PTFE) tube with an outer diameter of 2.3 mm, an inner diameter of 1.5 mm, and a length of 2.2 m. The PTFE tube encloses a stainless steel wire electrode (*active electrode*) of 0.1 mm diameter and a tungsten spike at the distal end of the electrode. The electrode is powered by a high voltage electrosurgical generator. In our experiments the distance between the APC probe and the treated biological tissue (*porcine kidney without capsule, defrosted at room temperature*) amounts to 2 mm. The biological tissue is treated when high voltage at the electrode of the argon plasma coagulator is applied, and the plasma is ignited that transmits electric current to the biological object [25], [26]. Thereby, surgical tissue-effects, like desiccation and coagulation [27], occurs.

Argon is used as the working gas for plasma ignition, which flows through the PTFE tube. In our experiment, nitrogen (5%) is admixed to the argon and is used as sensor-gas for OES diagnostics. Gas flow is 0.6 slm. The high voltage of both polarities is generated by the electrosurgical generator (see figure 8) with a trigger frequency of 350 kHz. During the positive phase of the high voltage, a spark discharge is ignited (see figure 9). Whereas, a normal glow discharge with typical negative glow near the cathode surface fills the gap between APC spike and tissue surface during the negative phase of the high voltage. By ignition of the spark discharge, a streamer crosses the gap between the APC spike and the treated tissue, and produces a thin conductive channel with a height of 2 mm (see figure 9a). During ionisation wave in direction to the anode, the gas in the plasma channel is heated and the plasma channel expands (see figure 9b). From current trace, the duration of the spark discharge amounts to 460 ns corresponding to the full width at half maximum of the current amplitude (see figure 8).

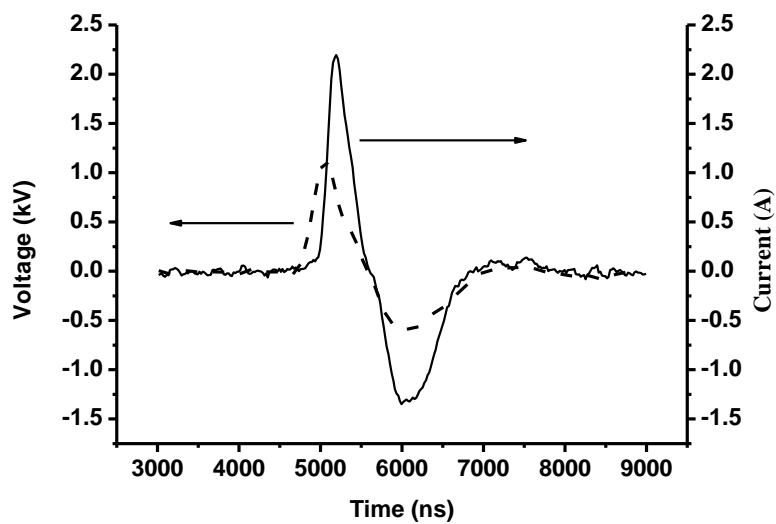

**Figure 8.** The voltage and current measurement at the electrode of the APC probe and the tissue on the opposite electrode for one voltage pulse.

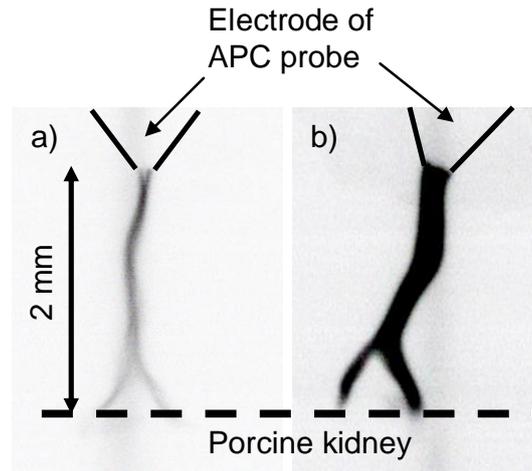

**Figure 9.** Inverse microphotograph of the APC discharge during the positive high voltage half period (a) streamer channel and b) spark discharge) on porcine kidney.

For characterization of the plasma parameters for the APC discharge OES diagnostic is used, where only emission intensities in the 'spark'-phase (positive phase) of the discharge is considered. The intensity of nitrogen emission in the 'glow'- phase (negative phase) of the discharge is an order of magnitude lower and is not characterized within the frame of this paper.

To calculate the steady state density of nitrogen molecular ions required for determination of electron density applying (16), we apply the reaction scheme presented in table 1.

**Table 1.** Plasma-chemical reactions and their rate coefficients for determination of nitrogen molecular ions density in APC discharge

| Reaction | Rate constant | Reference |
| --- | --- | --- |
| $Ar + e \rightarrow Ar^+ + 2e$ | Depends on $f_v(E)$ | Calculated applying eq. 12 |
| $Ar + e \rightarrow Ar^* + e$ | Depends on $f_v(E)$ | Calculated applying eq. 12 |
| $Ar^* + e \rightarrow Ar^+ + 2e$ | Depends on $f_v(E)$ | Calculated applying eq. 12 |
| $Ar^+ + 2Ar \rightarrow Ar_2^+ + Ar$ | $2.7 \cdot 10^{-43}$ m$^6 \cdot$s$^{-1}$ | [28] |
| $Ar_2^+ + e \rightarrow Ar + Ar$ | $8.7 \cdot 10^{-13}(T_e/300K)^{-0.67} \cdot (T_g/300K)^{-0.58}$ m$^3 \cdot$s$^{-1}$ | [28] |
| $N_2 + e \rightarrow N_2^+ + 2e$ | Depends on $f_v(E)$ | Calculated applying eq. 12 |
| $N_2^+ + e \rightarrow N + N$ | $1.75 \cdot 10^{-13}(T_e/300K)^{-0.30}$ m$^3 \cdot$s$^{-1}$ | [29] |
| $N_2(X) + e \rightarrow N_2(A) + e$ | Depends on $f_v(E)$ | Calculated applying eq. 12 |
| $N_2(A) + e \rightarrow N_2(B) + e$ | Depends on $f_v(E)$ | Calculated applying eq. 12 |
| $N_2(A) + e \rightarrow N_2(C) + e$ | Depends on $f_v(E)$ | Calculated applying eq. 12 |
| $N_2(A) + e \rightarrow N_2^+ + 2e$ | Depends on $f_v(E)$ | Calculated applying eq. 12 |
| $N_2^+ + N_2 + M \rightarrow N_4^+ + M$ | $5 \cdot 10^{-41}(300K/T_g)^{1.67}$ m$^6 \cdot$s$^{-1}$ | [30] |
| $N_4^+ + e \rightarrow$ products | $2.6 \cdot 10^{-12}$ m$^3 \cdot$s$^{-1}$ | [31] |
| $Ar^+ + N_2 \rightarrow N_2^+ + Ar$ | $1.1 \cdot 10^{-17}(T_g/300K)^{0.5}$ m$^3 \cdot$s$^{-1}$ | [32] |

Calculation under steady state conditions shows that argon atomic ions are produced mainly in stepwise ionisation via excited argon states (metastable and resonance) at low electric field conditions and direct ionisation at high electric field. Argon ions produce molecular ions $Ar_2^+$ in a three body reaction. Because of very effective molecular ion-electron recombination collisions, the density of $Ar_2^+$ ions amounts to about 0.2% of the atomic ions. Nitrogen molecular ions are produced mainly by a charge exchange reaction with the argon atomic ions. Lifetime of nitrogen molecular ions is short at high electron density because of the effective molecular ion-electron recombination. The steady state density of $N_4^+$ ions is negligible at our plasma conditions because of very effectively neutralisation process by collisions with electrons. Generally, the density of nitrogen molecular ions amounts to about 4% of argon atomic ions for our APC plasma conditions.

Electron densities calculated at APC discharge conditions by variable electric field applying (16-18) are presented in figure 10 (and with better resolution in figure 11). By determination of electron density using equation (16), a method of successive approximations is applied because nitrogen photoemission can be excited by reaction with excited argon atoms, whereby the rate constant $K_{N_2(C)}^{Ar^*}$ depends on the electron density. Four iterations are performed to receive a consistent solution. At that electron density determined after the fourth iteration differs from that determined by $K^{Ar^*}_{N_2(C)} = 0$ to about 13%.

The intersecting points of the curves presented in figure 10 at high electric field (about 1800 Td) give no reliable solution to the system of equations (16-18). This value of reduced electric field corresponds to a voltage of about 17 kV in the gas gap that is about an order of magnitude higher than the possible maximum value. Furthermore, the measured emission spectrum of nitrogen and current density cannot be interpreted in the range of high electric field. The intersecting points of the corresponding curves in the figure 10 in the high electric field region provide electron density values, which differ from each other to about two orders of magnitude. Hence, the points of intersection of the curves in the high field region are not suitable in the studied APC discharge condition.

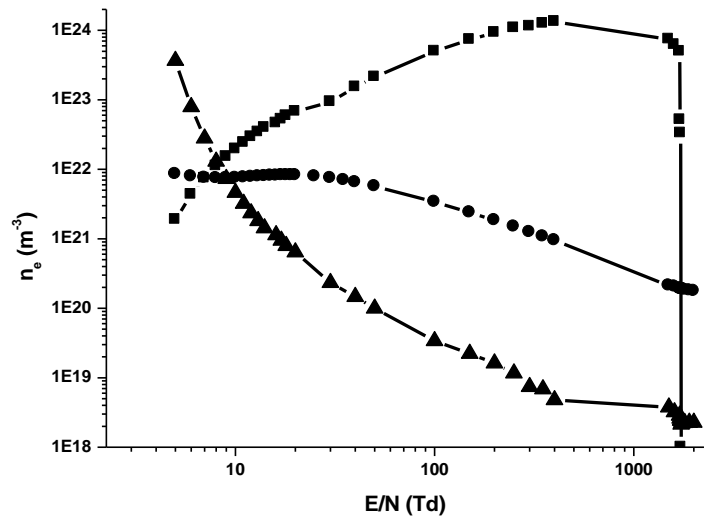

**Figure 10.** Graphical interpretation of equations system (16-18) for APC discharge in Ar/ $N_2$ mixture. ■ - applying ratio of nitrogen emissions (16), ▲ - absolute intensity of $N_2$(C-B,0-0) emission (17), ● - current density (18).

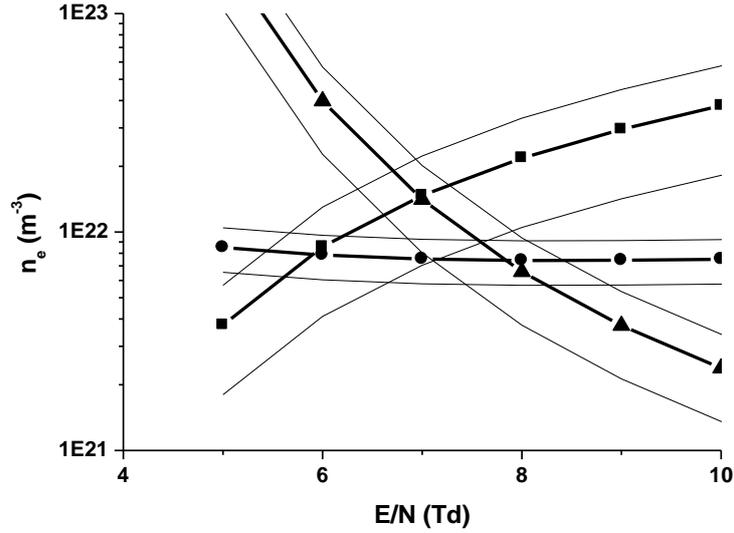

**Figure 11.** Graphical interpretation of equations system (16-18) for APC discharge in Ar/ $N_2$ mixture. ■ - applying ratio of nitrogen emissions (16), ▲ - absolute intensity of $N_2$(C-B,0-0) emission (17), ● - current density (18). Additional lines present confidence intervals of determined electron density.

A reliable solution to the system of equations (16-18) is at the low voltage region, where there is no "single point" of intersection but three different crossing points near the others can be found. Overlapping of all three confidence interval bands occurs in figure 10 near the point E/N = 6.9 Td and $n_e = 9.8 \cdot 10^{21}$ m$^{-3}$. Confidence intervals for the data presented in figure 11 are mainly caused by dispersion of measured intensities of nitrogen emissions $I_{N_2(C-B,0-0)}$ and $I_{N_2^+(B-X,0-0)}$ to about 25%, rate constants of emission quenching ($k_{N_2(C)}$, $k_{N_2^+(B)}$) to about 30% [10], and current measurement to about 4%.

Dispersion of measured intensities of nitrogen photoemission in APC discharge is abnormal large ($\frac{\Delta I}{I} = 0.25$). Possible reason of this effect could be the variation of the tissue impedance because of desiccation of the porcine kidney during the APC discharge application.

## 5. Discussion

The different discharge parameters measured for the homogeneous and filamentary DBD modes, and the APC discharge are presented in Table 2. The plasma parameters determined in the frame of the plasma-chemical model, where nitrogen emissions are excited at these plasma conditions by direct and stepwise processes (16-18) are presented in the Table 3.

**Table 2.** The parameters measured in the homogeneous and filamentary DBD modes, and APC discharge (nitrogen emission is corrected to the quenching). $t_p$ – plasma duration, $d_p$ – plasma diameter, $I_{avr}$ – averaged electric current.

| Kind of plasma | $T_g$ (K) | $t_p$ (ns) | $d_p$ (m) | $I_{avr}$ (A) | $I_{N_2(C-B)}$ (photons·s$^{-1}$·m$^{-3}$) | $I_{N_2^+(B-X)}$ (photons·s$^{-1}$·m$^{-3}$) |
|---|---|---|---|---|---|---|
| Homogeneous DBD | 330 ± 10 | 18 | $1.0 \cdot 10^{-2}$ | 0.275 | $1.20 \cdot 10^{21}$ | $2.74 \cdot 10^{20}$ |
| Filamentary DBD | 530 ± 30 | 20 | $1.0 \cdot 10^{-4}$ | 0.220 | $1.62 \cdot 10^{23}$ | $3.52 \cdot 10^{21}$ |
| APC discharge (spark) | 1400 ± 100 | 460 ± 10 | $1.7 \cdot 10^{-4} \pm 2 \cdot 10^{-5}$ | 1.060 ± 0.04 | $3.36 \cdot 10^{27}$ | $1.98 \cdot 10^{28}$ |

**Table 3.** Plasma parameters determined in homogeneous and filamentary DBDs and APC discharge applying OES and current measurement using system of equation (16-18) (see text).

| | Plasma parameters determined from the intersection of curves | | | | | |
| | eq. (16) and eq. (17) | | eq. (16) and eq. (18) | | eq. (17) and eq. (18) | |
| Kind of plasma | E/N (Td) | $n_e$ (m$^{-3}$) | E/N (Td) | $n_e$ (m$^{-3}$) | E/N (Td) | $n_e$ (m$^{-3}$) |
|---|---|---|---|---|---|---|
| Homogeneous DBD | 230 | $7.8 \cdot 10^{16}$ | 230 | $1.0 \cdot 10^{17}$ | 192 | $1.2 \cdot 10^{17}$ |
| Filamentary DBD | 196 | $8.7 \cdot 10^{20}$ | 196 | $9.1 \cdot 10^{20}$ | 192 | $9.3 \cdot 10^{20}$ |
| APC discharge | 7.0 | $1.4 \cdot 10^{22}$ | 5.9 | $7.9 \cdot 10^{21}$ | 7.8 | $7.4 \cdot 10^{21}$ |

The electric field determined in homogenous DBD is slightly higher than the electric field for ignition in our experimental conditions determined using Paschen law (about 180 Td) [33]. The homogeneous DBD is ignited in our experiment applying about 10 kV. This voltage is divided into three capacitances in series, namely ceramic with thickness of 1 mm, air gap of 2 mm, and glass plate with thickness of 1 mm. The capacitance of the air gap is lower than other capacitances in this electric circuit and the voltage in the air gap amounts to 90% of the applied voltage, for example 9 kV at plasma ignition. This value agrees well with the ignition voltage in air obtained for our experimental conditions applying Paschen curve [33]. The frequency of the high voltage pulses is low (300 Hz) and the residual charge in the air gap and on the electrode surfaces is also low. Therefore the electric field, which is not disturbed by space charge amounts to about 200 Td at plasma ignition. The electron avalanches, which start at discharge ignition, reach the opposite electrode in about 5 ns. After ionisation of the gas between the electrodes, the charge in the gap is polarised and a cathode fall is formed. The measured duration of plasma current in homogeneous DBD amounts to 18 ns which is long enough for formation of the cathode fall. The electric field and the photoemission intensity near the cathode fall is higher than those in other parts of the air gap. In our experiment, the spectrometer observes the whole gap during the homogeneous DBD. Maybe, these space variations of electric field and photoemission in homogeneous DBD are reasons of slightly higher than expected electric field (up to 230 Td) determined in our experiment.

The averaged electric field measured in our DBD operated in filamentary mode (about 195 Td) differs strongly from electric field determined in the streamer head propagated in air at atmospheric pressure (higher than 500 Td) [21], [34]. At our experimental conditions, streamer discharge that crosses the air gap initiates filamentary discharge on the surface of the ceramic [1]. The average electric field in discharge at the surface is higher than in the filament that crosses the air gap. To exclude influence of surface discharge on OES diagnostic, we use a diaphragm and thereby the spectrometer collects the photoemission of the middle part (about 1.5 mm) of the plasma filament ignited between Al-spike and ceramic. The streamer head crosses the observed volume with a velocity of about $5 \cdot 10^5$ m·s$^{-1}$, for example during 3 ns. Therefore, the spectrometer measures mainly photoemission of the streamer body. Electric field in this part of the streamer is lower than in the head [34]. Furthermore, after production of a conductive channel and the polarisation of charge, a cathode fall is formed and the electric field in the middle region between electrodes decreases.

The results presented in figure 7 and in table 3 testify that despite strong space and temporal variations of plasma parameters during ignition of transient discharges (DBD), the OES and current diagnostics provide consistent information about plasma parameters and can be used for averaged plasma characterization.

Our investigations show that, when plasma conditions are not clear *a priory*, both direct and stepwise excitation of nitrogen emissions by electron impact must be considered for OES diagnostics because equal intensities of nitrogen can be excited at low electric field with high electron density as well as at high electric field with low electron density [6]. To decide the reliable and appropriate solution at the studied plasma conditions, some additional information is required, for example, the difference in rotational temperature between the neutral nitrogen molecules and the nitrogen ions. Rotational distribution in emission spectra of neutral and ionized molecules was used for this purpose [6]. Diatomic molecular nitrogen ions are accelerated in the electric field and by collisions with neutral gas specie the rotational degrees of freedom are heated [35]. The rotational temperature of $N_2^+(X)$ increases and differs from the gas temperature, which corresponds to the rotational temperature of neutral molecules. The momentum of heavy nucleus is changed only slightly by electron impact excitation. Therefore, if $N_2^+(B-X)$ emission is excited by electron impact of nitrogen ions, the latter must have a higher rotational temperature than the surrounding gas. This temperature difference is used to conclude that stepwise excitation via $N_2^+(X)$ is important for excitation of ionic nitrogen emission $N_2^+(B-X,0-0)$ in a high frequency plasma in nitrogen at atmospheric pressure [6].

However, this method of the temperature difference is not always useful, e.g. by disturbance of nitrogen emission bands by intense photoemission of other molecules, like CN(B-X) in APC discharge.

As a solution to this problem, it is shown here that the measured current can be effectively used to identify the averaged plasma parameters existing at DBD and APC discharge conditions. Current measurement together with microphotography and calculation of drift velocity for variable electric field can complement OES diagnostics, and help to find reliable plasma parameters (see figure 7, 10) because only at this (reliable) condition the measured nitrogen emission and electric current can be described simultaneously. According to the results presented here, we conclude that the assumption of including only direct electron impact excitation of $N_2(X)$ state in the model for OES diagnostics of DBD conditions [1], [2], [3] is valid. This is valid not only in the homogeneous mode, where electron (and nitrogen ions) density is low, but also in the filamentary mode, where electron (and ions density) is three order of magnitude higher than the homogeneous mode. Because of rapid charging of dielectric surface and therefore, short duration time of DBD pulse, the applied spectrometer observes mainly the propagation of the electron avalanches (in the homogeneous mode), and streamer head (in the filamentary discharge). In both cases, ionization front propagates in the neutral gas with high ionisation potential and for that high electric field is required. At this condition, stepwise excitation of $N_2^+(B-X)$ emission via $N_2^+(X)$ is negligible in comparison to direct electron impact excitation.

The APC discharge is ignited by applying high voltage. While applying high voltage, the APC discharge consists of two different discharge types namely a spark and a glow discharge during the positive and the negative half period. For our investigations, the time-integrated measurements with the applied spectrometer represent photoemission mainly from the positive phase glow discharge. Intensity of photoemission in negative phase of applied voltage determined using microphotography amounts only 10% of total photoemission measured in the APC discharge. Hence, most photoemission is produced during positive phase of APC discharge and the averaged plasma parameters determined by applying OES diagnostic (16, 17) characterize spark and positive phase of glow discharge.

The exact solution for the system of equations (16-18) by graphical interpretation is presented by the intersection of all three curves in one point. But because of inaccuracy of the measured parameters, namely intensities of photoemission, current, diameter of the plasma channel, etc., three crossing points near each other are observed at all three discharge conditions (see figure 7, 11).

As was mentioned above, current measurement can complete OES diagnostics by characterization of microdischarges and help to determine reliable plasma parameters. Furthermore, combination of numerical simulation, current measurement, and ratio of nitrogen molecular emissions provide the possibility to obtain plasma parameters using equations (16) and (18) without using a calibrated spectrometer because the vibrational structure in $N_2(C-B)$ emission corresponds to Franck-Condon factors, is independent of plasma conditions and can be used for relative calibration itself. This possibility will be studied in our forthcoming experiments.

## 5. Summary

Three diagnostics methods namely optical emission spectroscopy, numerical simulation, and current measurements are applied for characterization of three transient discharges, namely filamentary and homogeneous DBDs in air, and APC discharge with very different plasma conditions. Nitrogen is used as sensor gas, where the emission of nitrogen molecule is studied and temporal and space averaged plasma parameters (electron distribution function and electron density) are determined. Two plasma-chemical models for excitation of nitrogen emission are considered, namely excitation by electron impact of nitrogen molecule in ground state and step wise excitation of nitrogen emission by electron impact excitation of nitrogen ions and metastables. It was established that the former can be applied for characterization of transient discharge, like DBD, while the latter is valid for high density discharges like spark that is ignited in APC source. It has been shown that current measurement is a consistent diagnostic method and can complement optical emission spectroscopy and numerical simulation.

**Acknowledgement**

This work is supported by 'Deutsche Forschungsgemeinschaft' (DFG) within the frame of the research group 'FOR1123 – Physics of Microplasmas' and ERBE Elektromedizin GmbH, Tuebingen.

# References

[1] Kuchenbecker M, Bibinov N, Kaemling A, Wandke D, Awakowicz P and Vioel W 2009 *J. Phys. D: Appl. Phys.* **42** 045212
[2] Rajasekaran P, Mertmann P, Bibinov N, Wandke D, Vioel W and Awakowicz 2009 *J. Phys. D. Appl. Phys.* **42** 225201
[3] Rajasekaran P, Mertmann P, Bibinov N, Wandke D, Vioel W and Awakowicz P 2010 *Plasma Processes and Polymers* **7** 665-75
[4] Kühn S, Bibinov N, Gesche R and Awakowicz P 2010 *Plasma Sources Sci. Technol.* **19** 015013 (8pp)
[5] Bibinov N, Knake N, Bahra H, Awakowicz P and Schulz-von der Gathen 2011 *J. Phys. D: Appl. Phys.* **44** 345204
[6] Rajasekaran P, Ruhrmann C, Bibinov N and Awakowicz P 2011 *J. Phys. D. Appl. Phys.* **44** 485205
[7] Farin G and Grund K E 1994 *End. Surg.* **2** 71
[8] Bibinov N, Halfmann H, Awakowicz P and Wiesemann K 2007 *Measurement Science and Technology* **18** 1327
[9] Pothiraja R, Bibinov N and Awakowicz P 2010 *J. Phys. D: Appl. Phys.* **43** 495201 (10pp)
[10] Pancheshnyi S V, Starikovskaia S M, Starikovskii A Y 2000 *Chem. Phys.* **262** 349-57
[11] Laux C O and Kruger H 1992 *J. Quant. Spectrosc. Radiat. Transfer* **48** 9-24
[12] Itikawa Y 2006 *J. Phys. Chem. Ref. Data* **35** 31
[13] Crandall D H, Kauppila W E, Phaneuf R, Taylor P O and Dunn G H 1974 *Phys.Rev.* **A9** 2545
[14] Dasgupta A, Blaha M and Giuliani J L 1999 *Phys. Rev.* **A61** 012703(10)
[15] Bacri J and Medani A 1982 *Physica* **112C** 101
[16] Armentrout P B, Tarr S M, Dori A and Freund R S 1981 *J. Chem. Phys.* **75** 2786
[17] Bibinov N, Rajasekaran P, Mertmann P, Wandke D, Vioel W and Awakowicz P 2011 *Biomedical Engineering, Trend in Material Science*, published by InTech, Croatia, edited by Anthony N Laskovski
[18] Code EEDF, available from Napartovich A P, Triniti Institute for Innovation and Fusion Research, Troizk, Moscow Region, Russia
[19] Herron J T 1999 *J. Phys. Chem. Ref. Data* **28** 1453
[20] Panousis E, Merbahi N, Clement F, Ricard A, Yousfi M, Papageorhiou L, Loiseau J-F, Eichwald O, Held B and Spyrou N 2009 *IEEE Trans. Plasma Sci.* **37** 1004
[21] Bonaventura Z, Bourdon A, Celestin S and Pasko V P 2011 *Plasma Sources Sci. Technol.* **20** 035012
[22] Rajasekaran P, Oplaender C, Hoffmeister D, Bibinov N, Sushek C V, Wandke D and Awakowicz P 2011 *Plasma Process. Polym.* **8** 246-55
[23] Lofthus A and Krupenie P H 1977 *J. Phys. Chem. Ref. Data* **6** 113
[24] Kossyi I A, Kostinsky A Y, Matveyev A A and Silakov V P 1992 *Plasma Sources Sci. Technol.* **1** 207-220
[25] Tjandra J J and Sengupta S 2001 *Dis Colon Rectum* **44** (12)
[26] Taieb S, Rolachon A, Cenni J-C, Nancey S, Bonvoisin S, Descos L, Fournet J, Gerard J-P and Flourie B 2001 *Dis Colon Rectum* **44** (12)
[27] Storek D, Grund K E, Schuetz A, Seifert H C, Farin G and Becker H D 1994 *Endoskopie heute* **2** 163
[28] Bogaerts A and Gijbels R 1999 *J. Appl. Phys.* **86** 4124
[29] Peterson J R, Le Padellec A, Danared H, Dunn G H, Larsson M, Larson A, Stromholm C, Rosen S, af Ugglas M and van der Zande W J 1998 *J. Chem. Phys.* **108** 1978
[30] van Koppen P A M, Jarrold M F and Bowers M T 1984 *J. Chem. Phys.* **81** 288-97
[31] Cao Y S and Johnsen R 1991 *J. Chem. Phys.* **95** 7356
[32] Anicich V G 1993 *J. Phys. Chem. Ref. Data* **22** 1469
[33] Raizer Y P 2002 *Gas Discharge Physics*, Springer, Berlin
[34] Pancheshnyi S, Nudnova M and Starikovskii A 2005 *Phys. Rev.* E **71** 016407
[35] Bibinov N, Dudek D, Awakowicz P and Engemann J 2007 *J. Phys. D: Applied Phys.* **40** 7372